# Predictive multiview embedding

## M LuValle

**Multiview embedding is a way to model strange attractors that takes advantage of the way measurements are often made in real chaotic systems, using multidimensional measurements to make up for a lack of long timeseries. Predictive multiview embedding adapts this approach to the problem of predicting new values, and provides a natural framework for combining multiple sources of information such as natural measurements and computer model runs for potentially improved prediction. Here, using 18 month ahead prediction of monthly averages, we show how predictive multiview embedding can be combined with simple statistical approaches to explore predictability of four climate variables by a GCM, build prediction bounds, explore the local manifold structure of the attractor, and show that even though the GCM does not predict a particular variable well, a hybrid model combining information from the GCM and empirical data predicts that variable significantly better than the purely empirical model.**

Since Lorenz first wrote about the chaos in the climate system [1]. The problem of predicting climate and weather has been connected to the general problem of predicting a partially unknown chaotic system. The climate problem also provides a canonical example of attempting to predict such systems, with the availability of both empirical models of real systems, and computational models developed to approximate the systems using first principles. In this paper we will use EdGCM[2,3] as our computational model and use runs from it based on injecting different levels of greenhouse gases and using the period where the temperatures are equilibrating as different tunings for a rising global surface temperature. Our empirical data will be precipitation and temperature data from 1960-2012 in volcano national park in Hawaii, data over the same time period of a normalized temperature difference between 2 bodies of water, and the multivariate ENSO index[4]. Our goal will be to understand how best to predict the last 55 months of the real data using a combination of empirical and computer generated data. To do this we need to understand more about predictive multiview embedding.

Multiview embedding [5] combines multiple different embeddings [6-8] to develop a diffeormorphic (embedding is 1-1, onto, and preserves derivatives ) picture of the strange attractor of a chaotic dynamic system. Predictive multiview embedding takes advantage of the theory underlying the embeddings to provide a picture not only of the dynamic system but of some of the uncertainties surrounding prediction in such systems.

A video developed by the Sukihara Laboratory (*http://tinyurl.com/EDM-intro*) provides an excellent introduction of how embedding works. With predictive multiview embedding, one measurement becomes special, it is chosen to be the one foremost in time. In certain conditions the embedding is diffeormorphic with the original attractor [7] but even if those conditions don't obtain, under a generalization of the chaoticity hypothesis[9] , each embedding provides an average prediction across several manifolds and can still be effectively predictive. Further, the residuals from an average prediction provides some insight into the manifolds being averaged over. The manifolds themselves are somewhat curious. The flow (and hence measurement) is attracted to "the expanding manifolds". In computer models this results in near confinement of the flow but in reality with open dynamic systems the expanding manifolds are idealizations as external perturbations constantly kick the system slightly off.

Embedding was first called out in [20] then extended and made more rigorous in [6-8]. The idea of embedding, particularly multiview embedding is the underlying concept for the approach to estimation taken here, and the notion of a delay map underlies embedding. To give an example of predictive multiview embedding we take a set of explanatory variables, say temperature at a given weather station at time point 18 months and, and 24 months before current time (-18, and -24), Multivariate Enso index[10] at time points -20, -22, and -26, and a linear combination of temperatures over two regions of the pacific ocean at times -18,-19, and -25.  Call this the X vector at time 0.  Add to that precipitation at the same weather station referred to above at time stamp 0, which we call Y at time 0. This characterizes the basic vector for delay map for predictive embedding 18 months ahead. A *delay map* is then constructed by shifting the whole vector (X,Y) into the past repeatedly for a large number of time steps.

The embedding theory [7] states that the set of sufficiently high dimensional multivariate timeseries of a chaotic system that are diffeomorphic to the original attractor are "prevalent" in the collection of such timeseries (delay maps), which in the space of infinite dimensional functions is a good substitute for having probability 1 [11]. Consider then the construction of several such embeddings each with the sufficiently high dimension and with the special choice of one term, a term at least k units of time in the future of all the other measurements, and corresponding to a particular measurement (e.g temperature or precipitation at a particular geographic point averaged over a given time period, or numerical value of an empirical function [4]). Almost all of these sets are diffeomorphic to the original attractor and therefore to each other, so for a point to be predicted there is a tangent plane. Each embedded point has such a tangent plane, and an asymptotically consistent nearest neighbor linear model for the future most term in each of these embeddings will provide regression estimates for the linear terms in Taylor series expansion for the embedding manifold. The prediction is the application of these terms to the point to be predicted. The Taylor series are different for each different embedding, but asymptotically they predict the same point if they are nearest neighbor predictions for the same point, and the expanding manifolds of the attractor are sufficiently separated. Bickel and Li[12] and Levina and Bickel [13] together provide an asymptotically optimal method of estimating single manifolds from data. Here we use a slight modification of their method, fitting the linear model at each point using least angle regression [14] to choose the prediction based on Mallow's Cp [15].  Predictability is evaluated using predictive correlation, the correlation between predicted values and observed values.

If the dimension of the embedding is too small, then the manifold that is being estimated may actually be an average of multiple manifolds (this statement can be made rigorous under the chaoticity hypothesis).  Similarly if the time is far ahead, or the neighborhood covered by the nearest neighbor estimate is large enough, multiple manifolds will also be included. This will show up in an estimate of the residuals for each prediction, as multiple modes in the residuals from the distribution. The distribution of single nearest neighbor predictions based on each embedding can be reconstructed and plotted against the multiview prediction (based on an average of the k nearest neighbor regressions for each embedding) to provide an asymptotically correct prediction bound (asymptotically correct by the ergodicity of the attractor and the asymptotic contraction of the nearest neighbor neighborhood). In the case of measurement error in the physical measurements being used in the embedding, there may not be an escape from estimating an average of several manifolds. While this is a bit of a difficulty, this automatically implies that each multidimensional measurement has a ball of uncertainty around it. Since the prediction procedure depends on nearest neighbors in a training sample, once those nearest neighbors are within the measurement error radius, predictions of sequential measurements acquire an asymptotic independence that allows derivation of rules for statistical inference.

In parallel to the development of embedding, work on the ergodicity of the chaotic dynamics[16,17] showed that under certain conditions chaotic dynamics results in an ergodic limit, which for special cases induces an ergodic measure which is absolutely continuous with respect to Lebesgue measure. The Chaoticity hypothesis [10] and various generalizations [18] indicate that it might be reasonable to apply many of these results to measurements on chaotic physical dynamic systems. The independent laws of large numbers for the chaotic system, and for the sampling for the multiview embedding provides a basis for the predictive bounds.  In particular the distribution of the single nearest neighbors is a predictive distribution, while the average of the linear predictions estimates the average manifold.

The differential geometry of chaotic systems [15] provides interesting information when thinking about comparing real data vs model generated data. In particular, at each point the attractor can be decomposed into orthogonal sets of manifolds, some expanding, some contracting. Whereas the computer model is almost perfectly following the expanding manifolds (for the dynamics represented in the computer model) in the real world system there is interference arising from exogenous systems. In climate and weather modeling this can include anything from the impact of solar flares to deforestation effects. This can lead to a situation where even if the computer model has systematic errors, it might provide useful information for predicting the real climate. For example the fluctuation dissipation theorem states that the natural fluctuations of the system follow the same linear response through time as small perturbations to the system will. The artificial climate system may give very precise estimates of not quite correct response, while the real climate system may give extremely noisy estimates of the correct response. If the not quite correct response has small enough bias, it can be used to improve prediction.

For prediction, we must be concerned with the uncertainty of the prediction process, so statistical inference becomes a necessity. Now we will explore the following approaches to inference, and then apply them to predicting the four variables in the weather system  the Multivariate ENSO index , local Precipitation in volcano national park. Local temperature in volcano national park, and a difference in temperature between two regions of the ocean.

i)   Topological englobement: This uses a measure of predictability to assess whether a class of models (e.g. computer models) represents a dynamic system well enough to be used in direct prediction of a particular class of measurement.

ii)   Residual based predictive bounds: this provides a way both of rescaling any inherent scaling issues in modeling, and of providing asymptotically correct prediction bounds.

iii)   Testing embedding variables for improving predictability: This method allows comparison of predictability with inclusion of different variables for multiview embedding. In particular here we will use it to see if the projection of real data onto a model which by itself is not good for prediction, can be used to improve prediction based on purely empirical predictive multiview embedding.

The idea of topological englobement is to test if a computer model of a dynamic system can in fact provide commensurate predictability for the real system. So consider a computer model with tuning parameters, $\theta$. If $\theta_0$ is the value of those tuning parameters for the real system, and assume the system is characterized by a system of ordinary differential equations. Take a set of $\theta_i$ as tuning parameters of a number of models, where $\theta_i$ lie in a very small ball closed centered at $\theta_0$ , then simulations based on those $\theta_i$ should spin out attractors very similar to one based on $\theta_0$. By the smoothness, the predictive correlations built using model i for training to predict model j should be close to those using model i to predict model 0. And in particular, using model prediction using model i to predict model 0 should give a very close predictive correlation to using model i to predict the other model j's. So predictive correlations using the models to predict real data should not produce a significantly different population from using models to predict models. In our study we use temperature change as the tuning parameter, for a GCM which clearly should not be able to predict these terms very well (to start with the areas averaged over for temperature and precipitation are 8 by 10 degrees, instead of at a single weather station).

Note that under the assumptions here the statistical uncertainty is contained entirely within the variation to the tuning parameters around the actual tuning parameters of the physical model. Statistical independence is provided by the different simulations. So a simple test of location of the population of predictive correlations of the real system using training samples from the model runs vs the population of predictive correlation of the samples from 1 run vs a different runs provides a way to falsify the hypothesis that the computational model represents the real system.

Figures 1-4 compare distributions of simulations predicting simulation ($\theta_i$ predicting $\theta_j$ in the above discussion), simulation predicting real data ($\theta_i$ predicting $\theta_0$ in the above discussion), real training data predicting real test data, and real training data augmented with the simulated data predicting the real data, to make a prediction of the real data, moving from left to right within each figure.

Figure 1: 18 month ahead predictability of difference of ocean pool temperatures

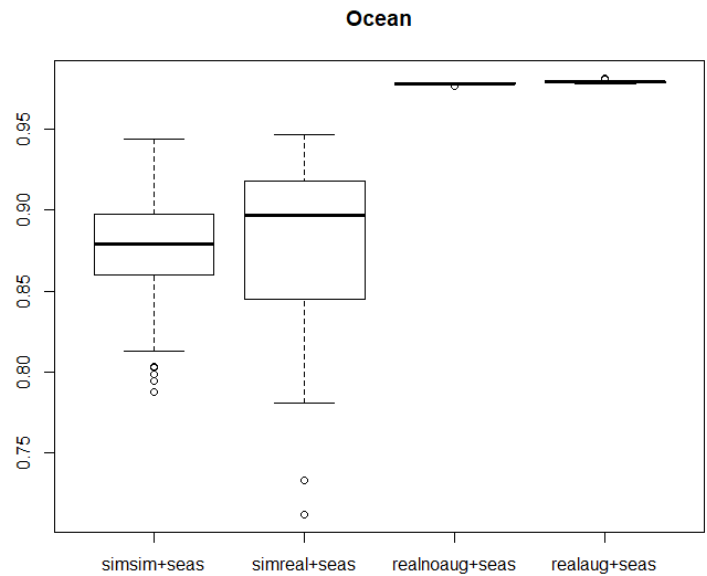

Figure 2: 18 month ahead predictability of precipitation at weather station

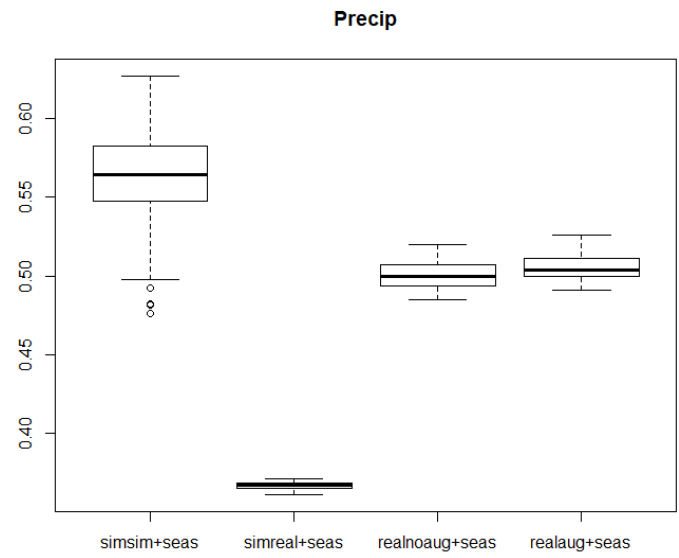

Figure 3: 18 month ahead predictability of air temperature at weather station

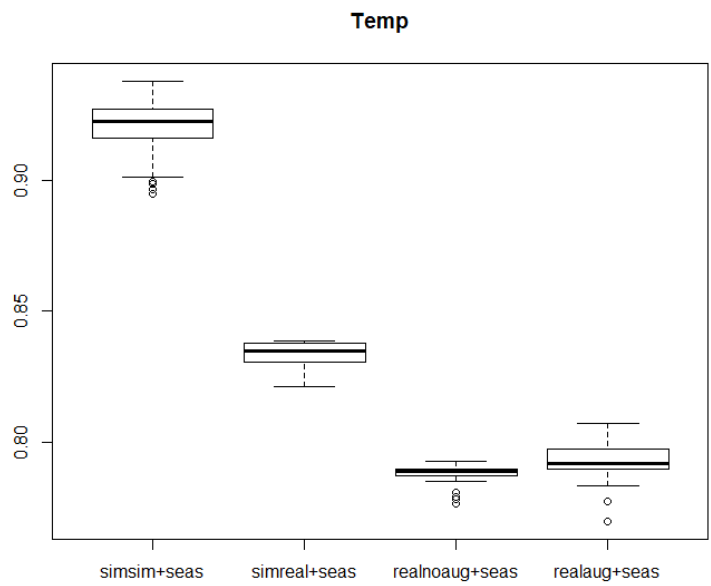

Figure 4: 18 month ahead predictability of Multivariate Enso Index

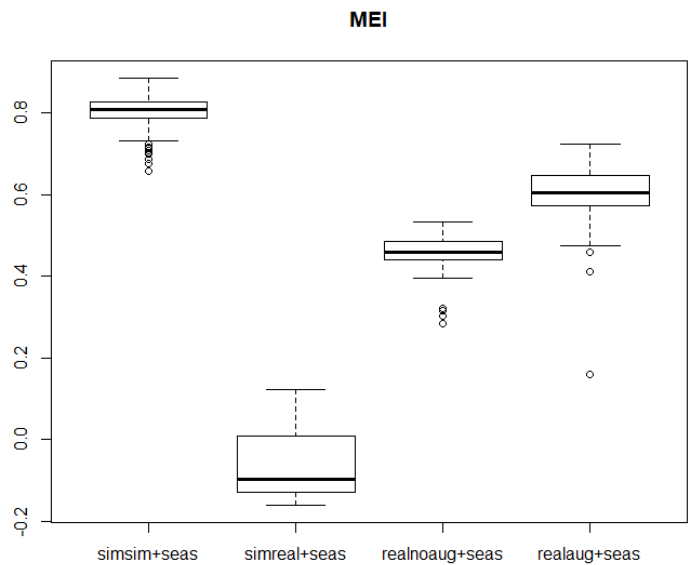

We see that the difference between the temperature of the two pools in the ocean are apparently as well modeled between computer models, as the computer models model the real world, though we do see the real world does provide a better model for itself than the computer models by themselves. Finally the augmented real model (using the projection of the real data onto the climate model data as another variable in the embedding) does even better.

For weather station level precipitation (with much lower base line correlations) the simulation models reflect the real world far less well than they reflect each other. However the real world models itself much better than the climate models. The augmented model, although appearing to offer some improvement over the real world model by itself, but this can not be said to be statistically significant (this judgement requires the 3rd statistical method, based on an assumption of measurement error).

For the weather station level temperature, we can see that the simulation system while predicting the real world less well than simulations predict each other, it does predict the real world better than the real world predicts itself, indicating some interesting skill in the climate model. The augmentation again apparently helps the real world model.

Finally the climate models predict each others enso index well but appears to have 0 predictive ability for the multivariate enso index. Part of this may be explainable by the fact that the index calculated on the models is not exactly the MEI, but is instead a linear combination of sea surface temperature and sea level pressure described in the supplementary information. On the other hand the augmented modeling appears to improve the prediction. Thus the predictive multiview embedding provides a framework for improving prediction using information in climate models, even when the climate models can not explicitly provide a direct path to prediction.

The plots above are based on a Pearson correlation coefficient so represent potential predictability based on the modeling procedure, not actual prediction. The local linear predictions tend to be shrunken, but the prediction bounds allow us to get bounds and estimates of the actual possible values. In the next series of plots we look at the kernel density estimate of the predictive distribution (black) vs the Gaussian distribution for (red) for the nine most recent times for each variable, the last 3 plots are False Discovery Rate [19] plots for the P values for 3 tests (red points represent interesting data points at that False Discovery Rate (.05 here). The 1st is for replication of the multiview embedding with different random views for each time predicted. The second (center) is for a Durbin based modification of the Kolmogorov Smirnov test between the raw data and the gaussian fit for each time predicted. The 3rd (left most bottom plot) is an EM algorithm based maximum likelihood tests for a mixture of a few gaussian distributions vs the best fit single gaussian.

Using the same order as before we see the Ocean temperature model is not far from a single gaussian. The precipitation is far from a gaussian model, but a mixture of a few gaussian distributions is not a good alternative. Both temperature and the MEI show a strong of evidence that they are well modeled by mixtures of a few gaussians, suggestive that a few manifolds are present. It may be possible the precipitation may represent a case of many many manifolds.

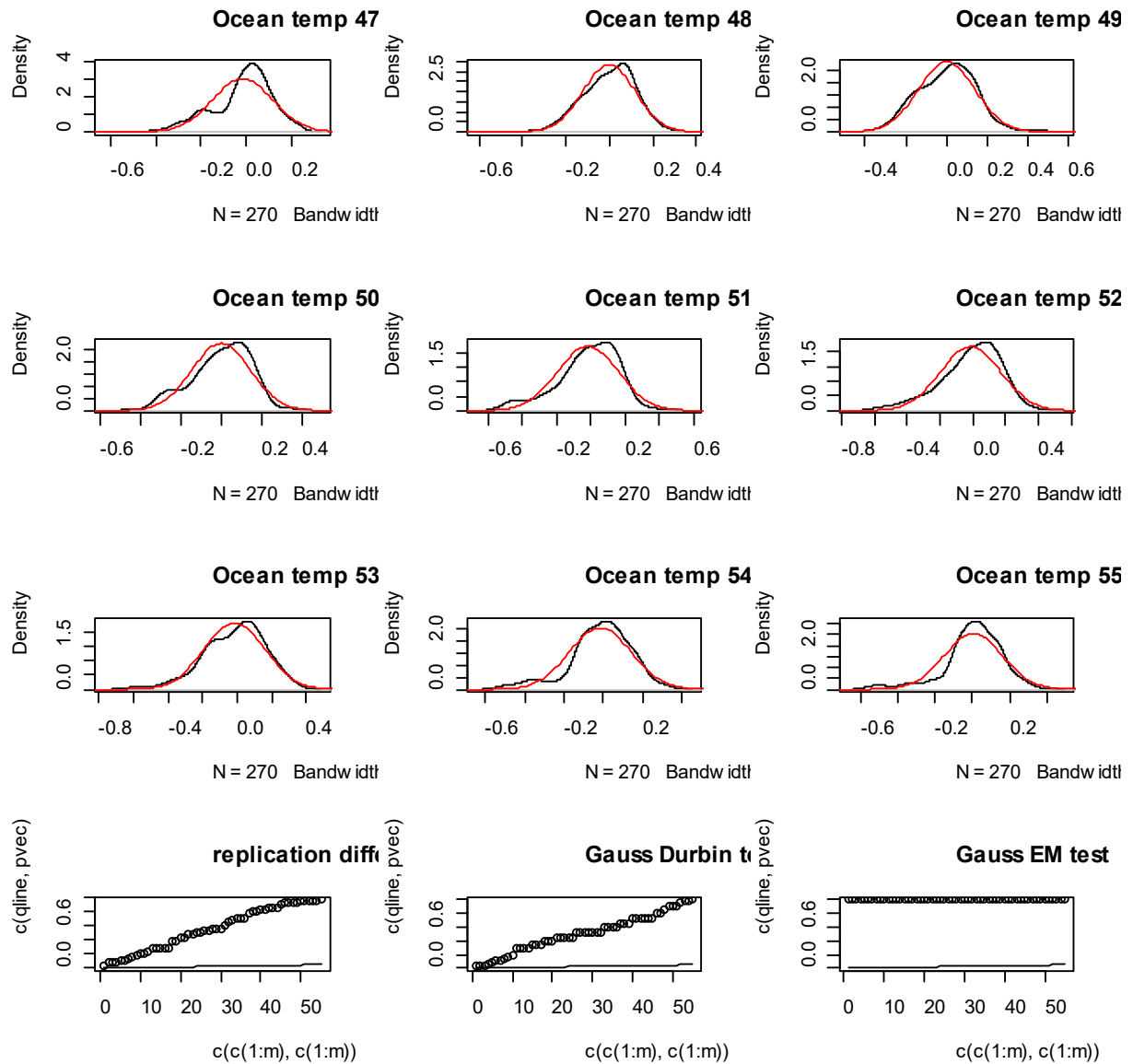

Fig 6

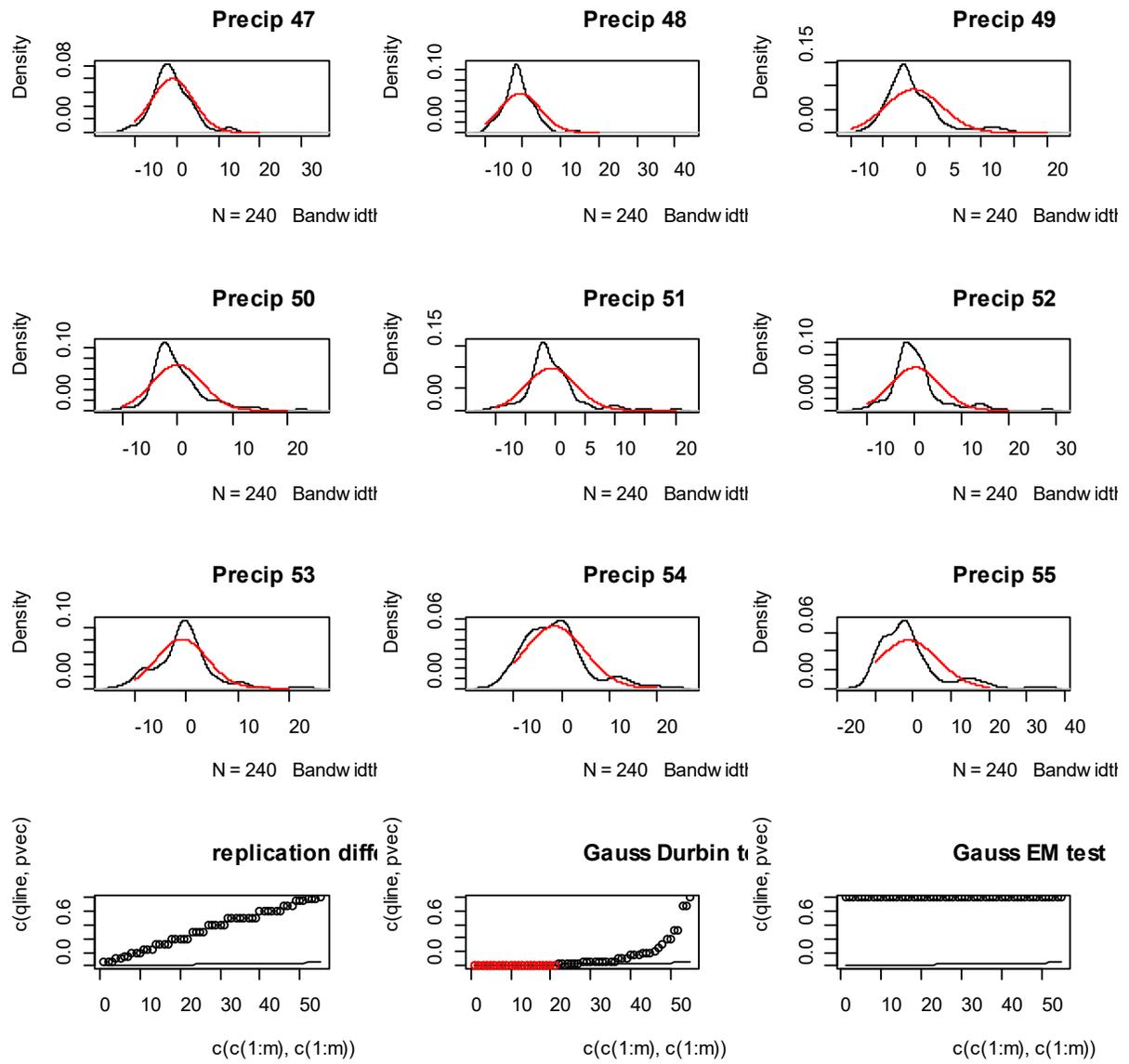

Fig 7

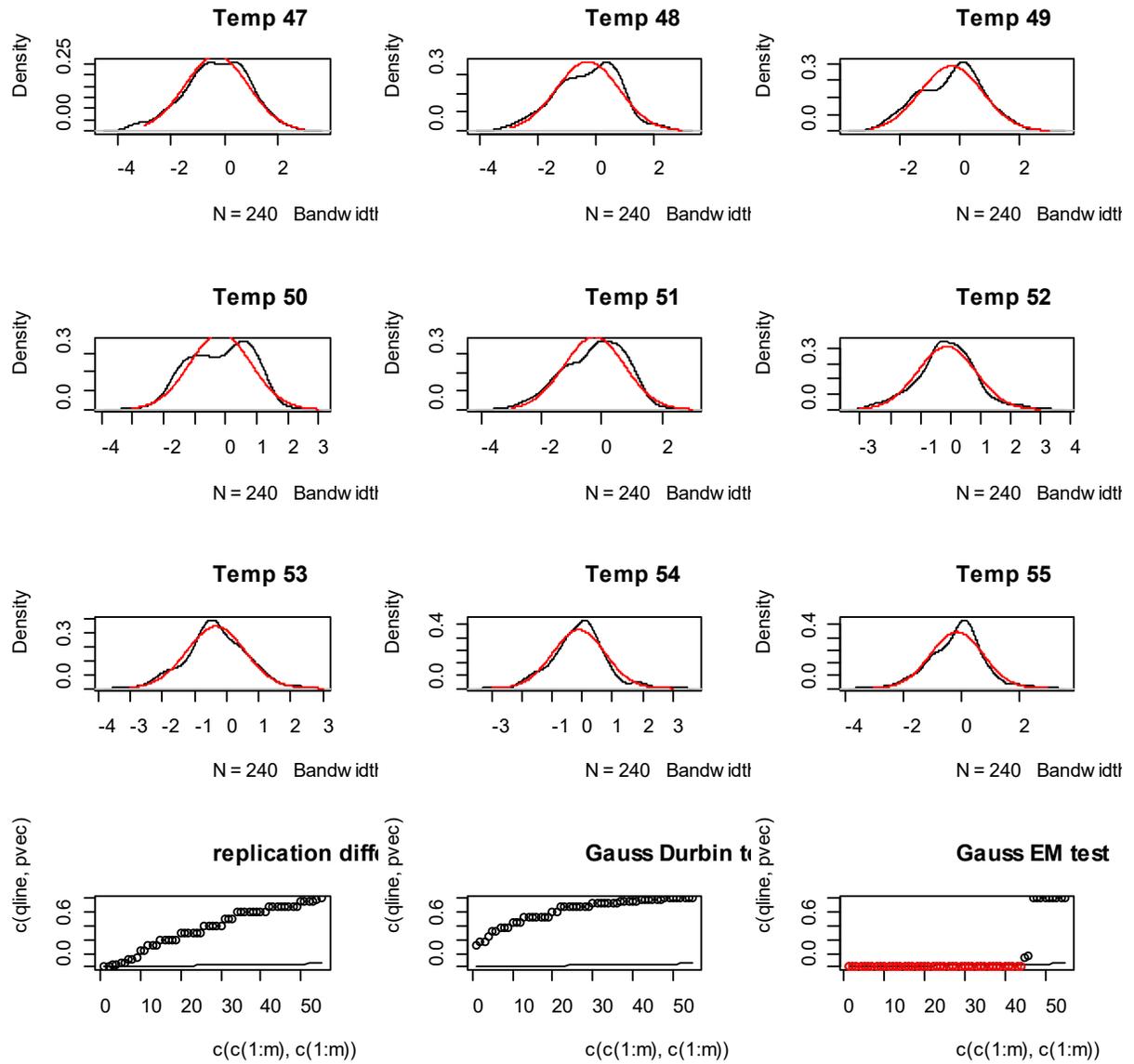

Figure 8

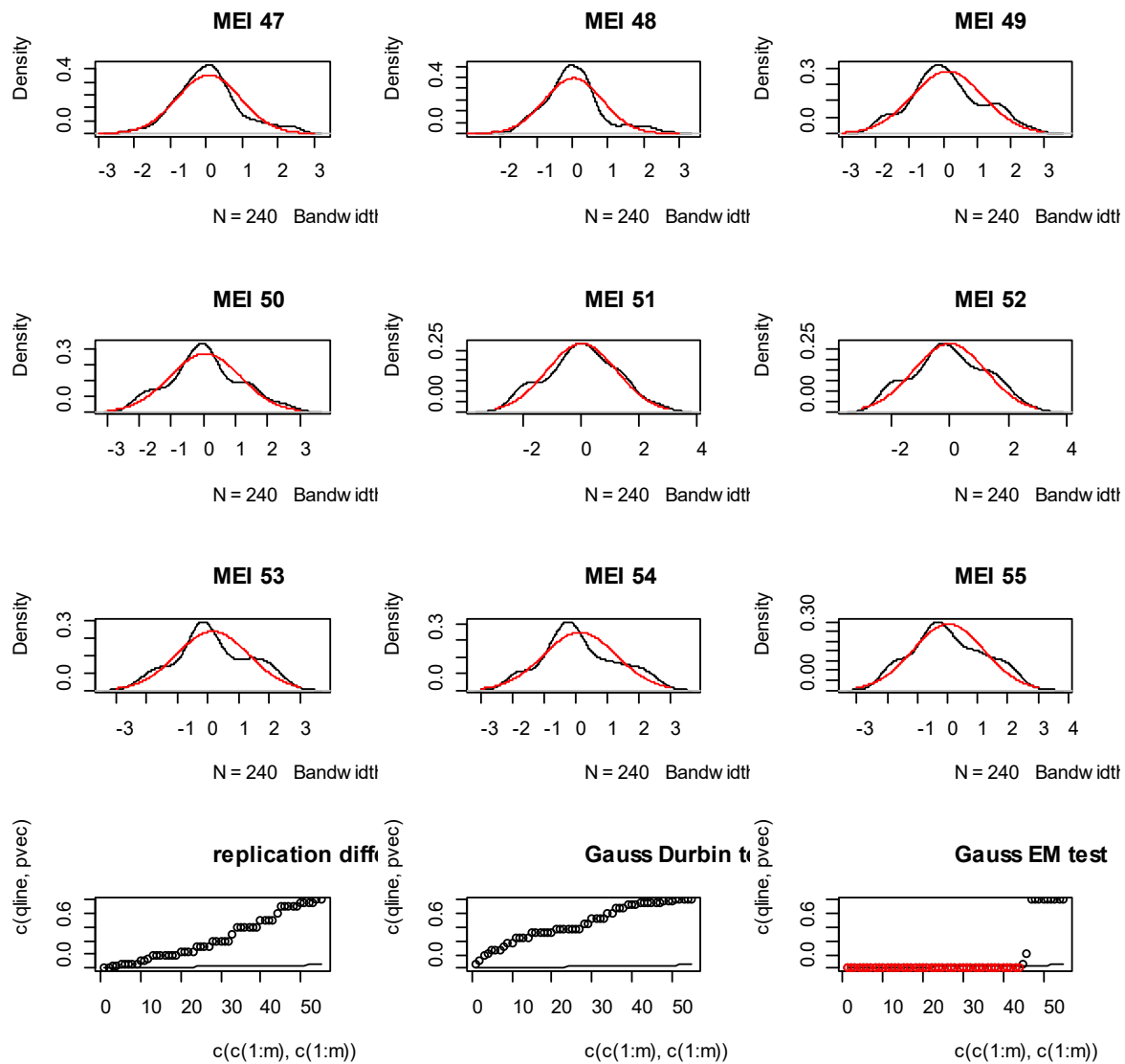

The problem of testing whether augmenting embeddings with projections onto climate model training data actually improves prediction is a more subtle statistical problem than the reasoning based on variation induced by the variations between $\theta_i$ in a neighborhood of $\theta_0$, or the ergodicity of the residuals induced by the natural ergodicity of the attractor. Predictions for a given time using a similar basis set of variables for the X variables will be correlated. Then because of the deterministic nature of the dynamics, subsequent times are strongly correlated with prior times. So any inference must be able to deal with any correlation across times. If there is no measurement error then it becomes necessary to separate the times more and more to get something close to asymptotic independence. However the nearest neighbor regression used for prediction plus assumed measurement error provides the framework for inference. Once the ball defined by the nearest neighbor terms is smaller than the typical measurement error, then while the regression is converging to the average manifold, any particular prediction for a time will have a distribution induced entirely by the deviation of the measurement error from the true value. Assuming independence here seems relatively innocuous. Thus we can do for example comparisons between two prediction approaches, by comparing differences pairwise at each prediction time, for example using a Wilcoxon test of pairwise differences. Applying this to the differences between the two rightmost distributions in the 1st set of 3 plots. We test if the embedding augmented by the projection onto the climate model provides better potential predictability than the embedding based purely on empirical data. For the difference in temperature between the ocean pools 18 months ahead, the pvalue for the test is ~.0003. For precipitation at the weather station in volcano national park it is .32, so not near significant. For temperature at the weather station it is ~.02, and for the MEI it is .0008. So for the two ocean measurements (18 months ahead) augmentation even by a climate model which is not particularly predictive by itself, the projection onto the climate model adds information to the embedding based prediction. There is some indication that it also helps in the temperature prediction but the noise is too high in the precipitation prediction to say it does. Part of the reason for this may be revealed in the residual plots for each. The MEI and temperature have clear distinct manifolds, the ocean pool difference may actually be represented well by a single manifold, but precipitation looks like there may be so many manifolds they are running into each other.

In conclusion predictive multiview embedding has been shown to be easily amenable to inference based on a number of approaches, two of them requiring nothing more than the chaoticity hypothesis, one requiring that the physical measurements of the real system be subject to some error, which is practically a tautology. Further it provides a simple way of exploring usefulness of low resolution climate models precision to improve predictability of important terms, such as the Multivariate enso index, which has shown to be correlated to a number of important climatic oscillations (e.g draught and fires in Australia and in North America). The methods described here are by no means optimal so significant work remains to improve on this however even as it is this method shows the way to a low cost approach for predicting important climatic variables.

There is also no reason that the same method cannot be applied to other dynamic phenomena, both for purely empirical modeling and to help in evaluating computational models of the phenomena.

<div align="center">References</div>